\documentclass{article}
\usepackage{graphicx} % Required for inserting images
\usepackage{subcaption} 
\usepackage{amsmath,amssymb,amsfonts} 
\usepackage{bbm,bm}
\usepackage{xcolor}
\usepackage{comment}
\usepackage{algorithm}
\usepackage{algpseudocode}
\usepackage{tikz}
\usetikzlibrary{shapes.geometric, arrows, positioning,  arrows.meta}
\usepackage{url,geometry}

\usepackage{natbib}
\bibpunct{(}{)}{;}{a}{}{,}

\usepackage{setspace}
\title{Uncertainty Estimation of the Optimal Decision with Application to Cure Process Optimization}
\author{
    Yezhuo Li\thanks{School of Mathematical and Statistical Sciences, Clemson University, Clemson, SC, USA} \and
    Qiong Zhang \footnotemark[1] \and
    Madhura Limaye\thanks{Manufacturing Science Division, Oak Ridge National Laboratory, Knoxville, TN, USA} \and
    Gang Li \thanks{Department of Mechanical Engineering, Clemson University, Clemson, SC, USA}
}
\date{~}

\begin{document}

\maketitle

\begin{abstract} 

Decision-making in manufacturing often involves optimizing key process parameters using data collected from simulation experiments. Gaussian processes are widely used to surrogate the underlying system and guide optimization.
Uncertainty often inherent in 
the decisions given by the surrogate model due to limited data and model assumptions. This paper proposes a surrogate model-based framework for estimating the uncertainty of optimal decisions and analyzing its sensitivity with respect to the objective function. %Using Gaussian process modeling, we generate a sample of optimal decisions by optimizing realizations from the posterior distribution of the surrogate model. This sample allows for the analysis of decision uncertainty and sensitivities to the objectives. 
The proposed approach is applied to the composite cure process simulation in manufacturing.

%This paper investigates uncertainty quantification in optimal decision-making for manufacturing processes, focusing on analyzing and modeling inherent variability. For some complex cases such as an expensive-evaluating or a black-box function, a Gaussian Process (GP) model is introduced to provide a probabilistic framework for quantifying uncertainty. Sensitivity analysis for optimal decisions identifies the parameters that most significantly influence optimal decisions, offering insights to guide future experiments and improve decision accuracy. Case studies illustrate the practical application of these methods, showcasing their capability to provide robust insights into uncertainty and its impact on optimal decision-making in the manufacturing process.
\end{abstract}


\newpage

\section{Introduction}

Decision-making is critical in many manufacturing processes, where optimizing key parameters is essential to achieving desired outcomes. 
However, some optimization problems in manufacturing lack closed-form expressions, making it difficult to apply gradient-based mathematical programming techniques. Instead, decisions are typically made by collecting data from simulation experiments and using surrogate statistical models to identify optimal settings.
For instance, in the composite cure process, it is important to investigate process parameters such as time and temperature of the cure cycle to minimize deformation in the final product \citep{limaye2025numericalsimulationinformedrapid}. The cure process can be simulated using finite element analysis in conjunction with the cure simulation tool. As noted in \cite{limaye2025numericalsimulationinformedrapid}, Bayesian optimization with Gaussian process is used to guide experiments and efficiently search for the optimal parameters in the cure cycle. 
Since the optimal decisions are driven by data, from practical perspective, it becomes important to assess the uncertainty associated with them, as well as to understand how this uncertainty affects the objective function values. 

Surrogate model, sensitivity analysis and optimization are common topics in statistics for computer experiments. 
Gaussian processes are widely used as surrogate statistical models for complex experimental systems \citep{sacks1989statistical,santner2003design}. Due to their probabilistic nature, Gaussian processes offer a convenient framework for quantifying prediction uncertainty across the entire input space. 
As a computational inexpensive surrogate of the real output function, Gaussian process 
can also be employed to estimate sensitivity indices, enabling the assessment of how uncertainty in each input contributes to the overall variability in the output function \citep{storlie2008multiple,marrel2009calculations,gramacy2020surrogates}.
Optimization is one of the important goals of developing simulation models (e.g., \cite{zhang2020sequential}).
Bayesian optimization is a widely used approach for solving those black-box optimization problems, where objective function evaluations are expensive and obtained through computational experiments (e.g., \cite{jones1998efficient, frazier2018tutorial}). By sequentially incorporating new experimental data, Bayesian optimization refines a surrogate model and proposes decisions that are optimal under the fitted model. However, the optimal decisions derived from this process are inherently subject to uncertainty—stemming from both limited data and the surrogate model itself. To the best of out knowledge, this aspect of decision uncertainty has not been fully discussed in the existing literature.

This paper aims to develop a surrogate model-based framework for estimating decision uncertainty and analyzing its sensitivity with respect to the objective function and apply the proposed methods to the application of cure process optimization. In this paper, we use Gaussian process as the surrogate model to fit data collected from simulation. We can collect a sample of the decisions by optimizing the realizations from the conditional Gaussian process given data. This sample can be used to construct the distribution of the optimal decision. Further more, we can perform sensitivity analysis based on this distribution.  
The proposed approach provides insights into the uncertainty associated with the optimal parameter settings in cure process simulations and quantifies the contribution of each input to this uncertainty, offering valuable guidance for practitioners in adjusting experimental settings in real-world scenarios.

This paper is organized as follows. Section \ref{sec:gpm} describes the proposed approach of surrogate model-based decision uncertainty estimation. 
%introduces GP and explores its application for estimating uncertainty. This section also extends to scenarios with constraints, highlighting how constrained optimization under GP models can enhance decision-making. To further assess the robustness of our approach. 
Section \ref{sec:sensitivity_analysis} illustrates how to use decision uncertainty in sensitivity analysis. %Using the distribution of optimal decisions, we derive sensitivity scores to identify which parameters most significantly affect optimal outcomes. This information helps guide future experiments and improve decision accuracy. 
Application of the proposed approach to cure process optimization is provided in Section \ref{sec:case_study}. %demonstrates the practical application of the proposed models and analysis methods. 
A conclusion is given in Section \ref{sec:conclusion}. %concludes the paper, summarizing key insights and outlining potential directions for future research.

\section{Surrogate Model-based Decision Uncertainty Estimation} \label{sec:gpm}

Consider an optimization problem 
\begin{equation}
\label{eq: opt}
  \bm x^\ast\in  \arg\min_{\substack{\bm x \in \mathcal{X}}} y\left( \bm x \right),
\end{equation}
where $\bm x=(x_1,\ldots, x_d)^\top\in\mathcal X\in \mathbb R^d$ is the $d$ dimensional decision variable and $y(\bm x)$ is an expensive-to-evaluate black-box objective.
This objective can be a function that characterizes a performance metric of a manufacturing process simulation under parameter setting $\bm x$. Our goal is to assess optimal decisions based on a limited budget of function evaluations and also to provide uncertainty estimation of the optimal decision. To this end, we propose a surrogate model-based decision uncertainty estimation approach.

Gaussian Process (GP) is a popular choice to surrogate black-box functions in manufacturing simulation \citep{williams2006gaussian}. 
%For any set of input points, the GP gives a multivariate Gaussian distribution over the corresponding outputs. 
Let 
%$\bm x^{\left(1\right)}, \cdots, \bm x^{\left(n\right)}$
$\mathcal{D} = \left\{ \bm x_1, \cdots, \bm x_n \right\}$
be a set of inputs in $\mathcal X$
and $\bm y  = \left( y \left( \bm x_1 \right), \cdots, y \left( \bm x_n \right)\right)^\top$ be a vector of corresponding outputs. 
Following \cite{jones1998efficient}, we model the outputs $y(\bm x)$ by a Gaussian process:
\begin{equation}
    Y \left( \bm x\right) = \mu + \epsilon \left( \bm x \right), \quad \epsilon \left( \bm x \right) \sim \mathcal{GP}\left( 0, \sigma^2 r(\cdot, \cdot) \right)
    \label{eq: model1}
\end{equation}
where $\mu$ is the constant mean of the Gaussian process, $\epsilon \left( \bm x\right)$ is a mean zero GP with 
variance $\sigma^2$ and correlation function $r(\cdot, \cdot)$. Therefore, for any two points $\bm x_i$ and $\bm x_j$, we have
\[
r(\bm x_i, \bm x_j)= \mathrm{Corr}\left[ \epsilon \left( \bm x_i \right), \epsilon \left( \bm x_j \right) \right].
\]
A commonly used correlation function is the Matérn family correlation function \citep{roustant2012dicekriging}. A Matérn $5/2$ correlation function is given by:
\begin{equation}
   r(\bm x_i, \bm x_j) = \left( 1 + \sqrt{5} r_{ij} + \frac{5}{3} r_{ij}^2 \right) \exp\left(-\sqrt{5} r_{ij} \right)
\label{eq:corr}
\end{equation}
with
\[
r_{ij} = \sqrt{\sum_{h=1}^{d} \frac{(x_{i,h} - x_{j,h})^2}{\ell_h^2}}
\]
measuring the distance between \( \bm x_i \) and \( \bm x_j \)
with unknown correlation parameters $\ell_1, \ldots, \ell_d$. 
In our numerical study, $\ell_h$'s are estimated by maximizing the likelihood function based on 
the GP assumption. 
%in \eqref{eq: model1}, profiling out the mean $\hat{\mu}$ and variance $\hat{\sigma}^2$ in \eqref{eq: y_mean&var}, and using a gradient‐based optimizer (L-BFGS-B) to search over $\ell_h$'s. 

Given the correlation parameters $\ell_h$'s, we can obtain the maximum likelihood estimators of $\mu$ and $\sigma^2$:
\begin{equation}
    \hat{\mu} = \frac{\bm 1^\top \bm R ^{-1} \bm y}{\bm 1^\top \bm R ^{-1} \bm 1}\quad\mathrm{and}\quad
    \hat{\sigma}^2 = \frac{\left(\bm y - \bm 1 \hat{\mu} \right)^\top \bm R ^{-1} \left(\bm y - \bm 1 \hat{\mu} \right)}{n}
    \label{eq: y_mean&var}
\end{equation}
where $\bm 1$ is a vector of ones with size $n$
and
$\bm R$ is an $n \times n$ matrix with the $\left( i, j \right)$-th entry $r \left(\bm x_i, \bm x_j \right)$. For any new input point $\bm x\in \mathcal X$, we denote
\[\bm r(\bm x)=
\begin{bmatrix}
r(\bm x,\bm x_1), 
r(\bm x,\bm x_2),
\ldots,
r(\bm x,\bm x_n)
\end{bmatrix}^\top.\] 
We obtain the conditional Gaussian process of $Y(\bm x)$ given $\bm y$ 
\begin{equation}\label{postGP:sigma}
 Y \left( \bm x \right) \mid \bm y\sim \mathcal{GP} \left( \hat{y} \left( \bm x \right), \Sigma \left( \bm x, \bm x'\right) \right). 
 \end{equation}
where
\[
\hat{y} \left( \bm x \right) = \hat{\mu} +\bm r(\bm x)^\top \bm R^{-1} \left(\bm y- \bm 1 \hat{\mu} \right)
\]
is the conditional mean function, and \[
\Sigma(\bm x,\bm x^{\prime})
=
\sigma^{2}\!\Biggl[
\,r(\bm x,\bm x^{\prime})\;-\;
\bm r(\bm x)^{\!\top}\bm R^{-1}\bm r(\bm x^{\prime})
\;+\;
\frac{\Bigl(1-\mathbf 1^{\!\top}\bm R^{-1}\bm r(\bm x)\Bigr)
\Bigl(1-\mathbf 1^{\!\top}\bm R^{-1}\bm r(\bm x^{\prime})\Bigr)}
{\mathbf 1^{\!\top}\bm R^{-1}\mathbf 1} \Biggl]
\,
\] is the conditional covariance function. 

Based on this conditional Gaussian process, an estimation of the optimal decision $\bm x^\ast$ is given by minimizing the mean function $\hat y(\bm x)$
\begin{equation}\label{eq:optmean}
    \hat{\bm x}^\ast\in \arg \min_{\substack{\bm x \in \mathcal{X}}} \hat y\left( \bm x \right).
\end{equation}
A random sample of size $M$ for the optimal decision $\bm x^\ast$ can be generated by 
optimizing $M$ independent realization functions drawn from the conditional Gaussian process
$Y \left( \bm x \right) \mid \bm y$.  Let $\tilde y_{(i)}(\bm x)$ with $\bm x\in\mathcal X$ be the $i$-th realization of the conditional Gaussian process $Y \left( \bm x \right) \mid \bm y$. For $i=1, \ldots, M$, we obtain
\[
\bm x^\ast_{(i)}\in \arg \min_{\substack{\bm x \in \mathcal{X}}} \tilde y_{(i)}(\bm x)
\]
The resulting optimal point from each realization forms a random sample $\left\{\bm x^\ast_{(1)}, \ldots, \bm x^\ast_{(M)}\right\}$
for estimating the uncertainty of the optimal decision $\bm x^\ast$.

In our implementation, we evaluate the realization functions $\tilde y_{(i)}(\bm x)$ on a finite set of input values. First, 
we generate a random $d$-dimensional Latin Hypercube design of size $N$ with points $\mathcal X_{(i),N}=\{\bm x_{(i),1}, \ldots, \bm x_{(i), N}\}$
for the $i$-th realization. Second,
we obtain the mean vector $(\hat y(\bm x_{(i),1}), 
\ldots, \hat y(\bm x_{(i),N}))^\top$ and the $N\times N$ covariance matrix with $(jk)$-th entry $\Sigma(\bm x_{(i),j}, \bm x_{(i),k})$, and generate a random sample 
\[(\tilde y(\bm x_{(i),1}), 
\ldots, \tilde y(\bm x_{(i),N}))^\top\]
from the multivariate normal distribution with the above mean vector and covariance matrix. Finally, we obtain 
\[
\bm x^\ast_{(i),N}\in \arg \min_{\substack{\bm x \in \mathcal X_{(i),N}}} \tilde y_{(i)}(\bm x)
\]
and forms the random sample of the optimal decision:
\begin{equation}\label{eq:mcsample}
\left\{\bm x^\ast_{(1),N}, \ldots, \bm x^\ast_{(M),N}\right\},
\end{equation}
which is referred to as the decision uncertainty sample in this paper. Using this sample, we are able to obtain the empirical distribution of the optimal decisions under the uncertainty of the fitted Gaussian process model. If the optimization problem has a black-box constraint, 
we can also fit a Gaussian process surrogate for the constraint function and generate a sample from the resulting conditional distribution to assess the feasibility. Detailed steps are provided in the Appendix. 

\begin{figure}[htbp]
  \centering
  % first sub‐figure
  \begin{subfigure}[b]{0.55\textwidth}
    \centering
    \includegraphics[scale=0.14]{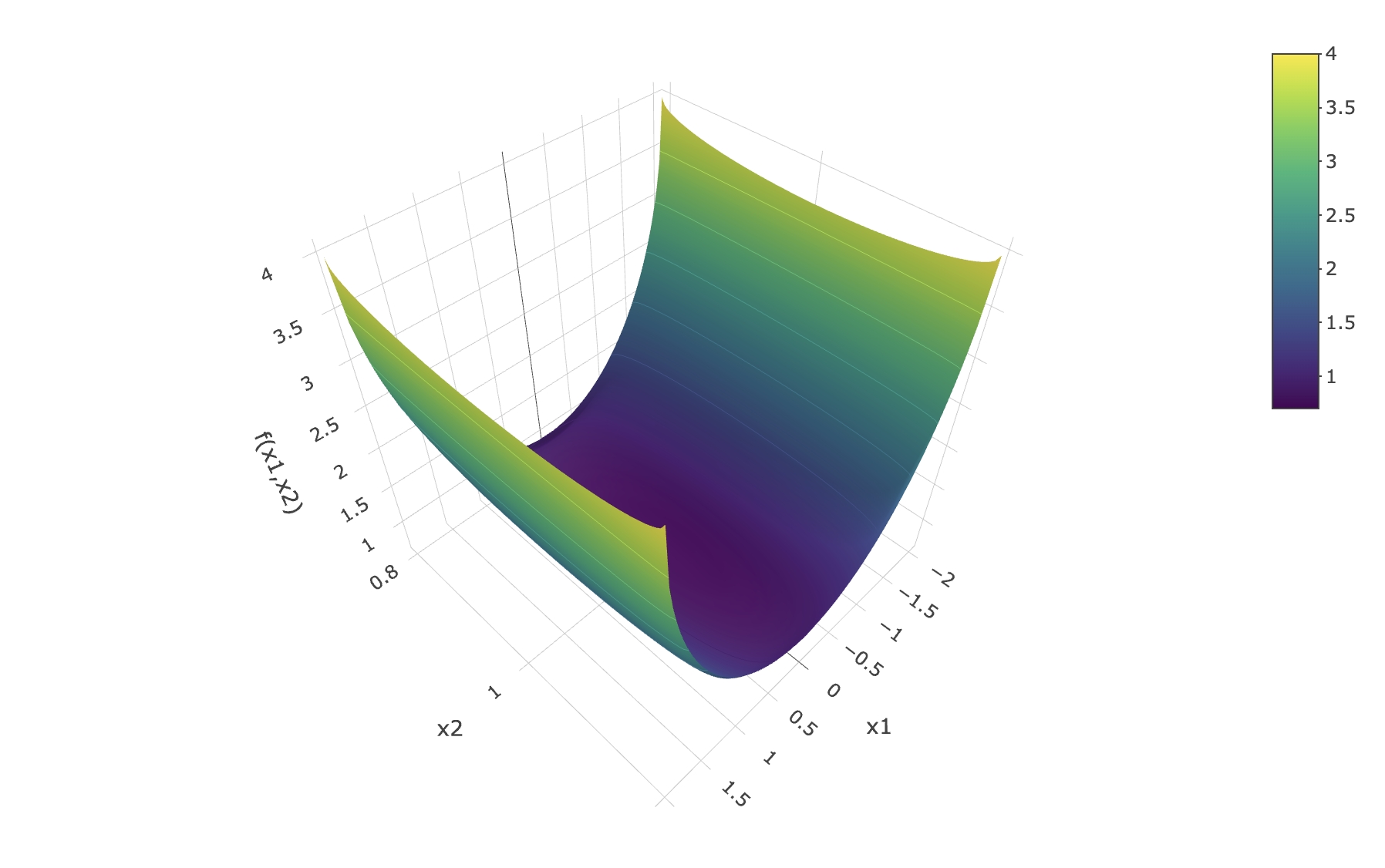}
    \caption{%Surface Plot of the Function in Equation \eqref{eq:2stage}
    }
    \label{fig:sub1}
  \end{subfigure}
  \hfill
  % second sub‐figure
  \begin{subfigure}[b]{0.44\textwidth}
    \centering
    \includegraphics[scale=0.11]{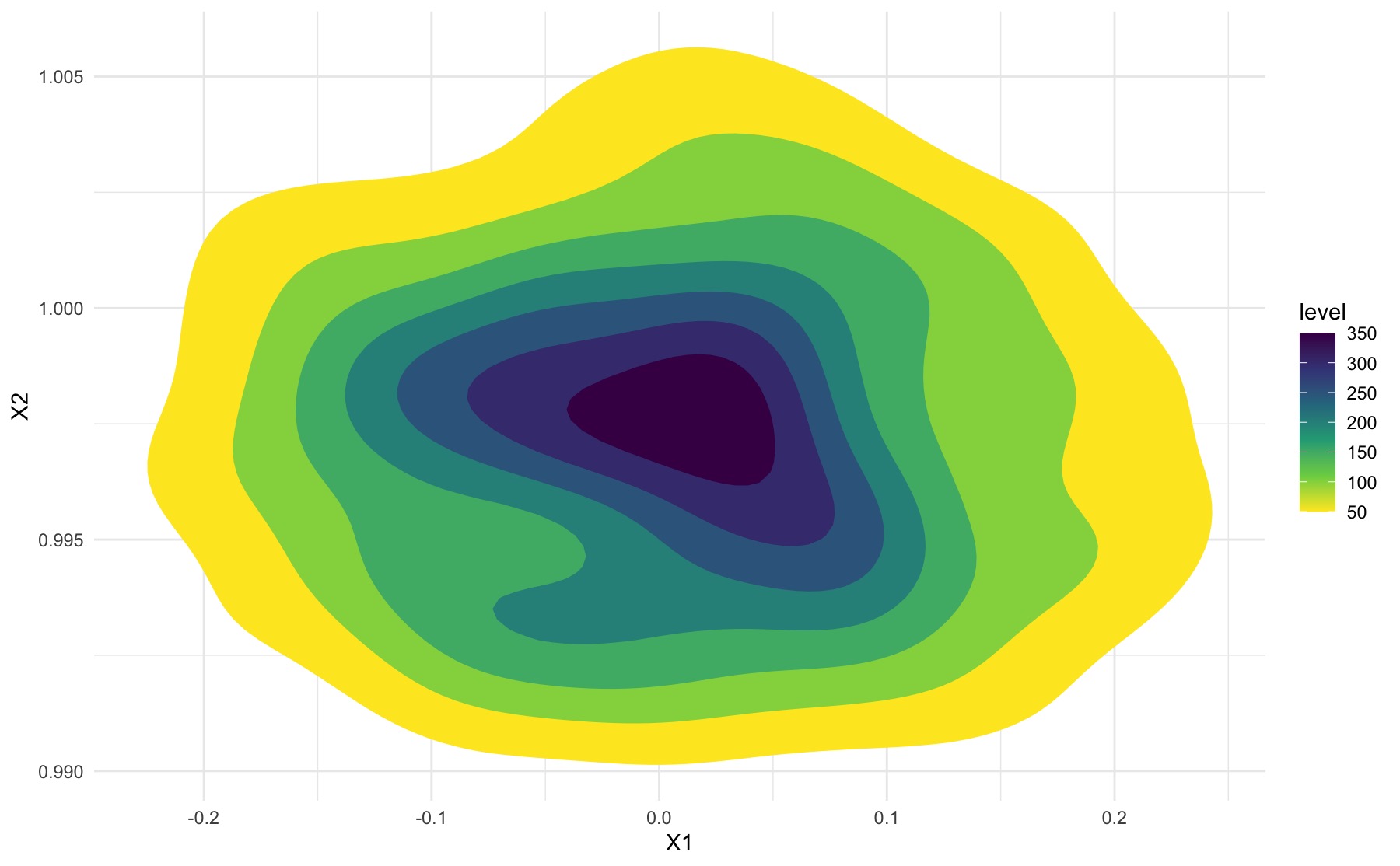}
    \caption{%Contour Plot of the Function in Equation \eqref{eq:2stage}
    }
    \label{fig:sub2}
  \end{subfigure}
  \caption{(a) Visualization of the function in \eqref{eq:2stage}; (b) the contour plot of the probability density function for
  the empirical distribution of \eqref{eq:mcsample}
for the optimization problem in \eqref{eq:2stage}.}
  \label{fig:Surface Plot}
\end{figure}

\paragraph{Illustration Example} We use an example to illustrate the proposed uncertainty estimation approach. Consider minimizing a 
two-dimensional function
\begin{equation}
    \min_{\substack{-2 \le x_1 \le 2\\ 0.8 \le x_2 \le 1.2}} y\left( x_1, x_2 \right) = -0.3 \sqrt{ 1- \frac{x_1^2}{4} } - 3 \sqrt{1- \frac{\left( x_2-1 \right)^2}{0.04}} + 4.
    \label{eq:2stage}
\end{equation}
The 3D visualization of this function \eqref{eq:2stage} is provided in Figure \ref{fig:sub1}, which
exhibits a semi-ellipsoidal geometry and possesses a unique global minimum at $\bm x^* = [0,\,1]^\top$. 

We use the proposed approach to assess uncertainties of 
the optimal decision based on a given dataset. First, we take $n=100$ runs of the Latin hypercube sample \citep{mckay1992latin} for the design set $\mathcal D$, and generate the corresponding outputs $\bm y$ using the function \eqref{eq:2stage} to fit a GP surrogate model. 
Let $N = 1000$ and $M = 500$, we can obtain a random sample of the optimal decision as in \eqref{eq:mcsample}. We depict the contour plot of the probability density function for the empirical distribution of this random sample.
This figure provides the visualization the uncertainty estimation of the optimal decision based on the surrogate GP model. 
In this Figure, the range of the optimal decision along the first coordinate is approximatedly from -0.2 to 0.2, whereas the range of the optimal decision along second coordinate is approximately from 0.990 to 1.005. Although the range of $x_2$ is smaller than $x_1$, it is not necessarily to indicate that the uncertainty contribution from $x_2$ is smaller than $x_1$. This motivates to use the empirical distribution of \eqref{eq:mcsample} to further perform sensitivity analysis to understand the sources of uncertainty contribution.   

\section{Sensitivity Analysis with Decision Uncertainty Estimation} \label{sec:sensitivity_analysis}
%Sensitivity analysis is applied for analyzing how the output of a model or system is affected by changes in its input parameters and identifying which inputs have the most influence on the output. 
%In this section, we perform a sensitivity analysis \cite{saltelli2008global} to evaluate how much impact each input variable has on the output.
%the differences between the effects of the original inputs and those of the estimated optimal solutions. 

The uncertainty estimation of the decision can be used to as the input distribution in sensitivity analysis. We focus on 
 Sobol' first indices and total sensitivity indices (e.g., \cite{gramacy2020surrogates}). Assume that the decision variable $\bm x=(x_1, \ldots, x_d)^\top$ follows a $d$-dimensional distribution with probability density function $p(x_1, \ldots, x_d)$. Let $x_{-i} = \{x_1,\dots,x_{i-1},x_{i+1},\dots,x_d\}$. The Sobol' first-order indices and the total sensitivity indices
 are given by
\begin{equation}\label{eq:index}
    S_{i} \;=\;\frac{\mathrm{Var}\!\bigl( \mathrm{E}[y(\bm x)\mid x_i]\bigr)}{\mathrm{Var}(y(\bm x))}\! \quad\mathrm{and}\quad
    S_{T_i}
\;=\;\frac{\mathrm{E}\bigl[\mathrm{Var}(y(\bm x)\mid x_{- i})\bigr]}{\mathrm{Var}(y(\bm x))},
\end{equation}
where the expectation and variance are taken with respect to the input distribution $p(x_1, \ldots, x_n)$ and the conditional distributions of $x_{-i}|x_{i}$ and $x_i|x_{-i}$, respectively.
The Sobol' first-order index $S_i$ represents the total output variance that is explained solely by varying $x_i$ on its own, where as $S_{T_i}$ is the total sensitivity index, measuring the total output variance explained by both the direct effect of $x_i$ and all interactions involving $x_i$. 
Both indices can be used to assess the contribution of 
$x_i$'s uncertainty to the output $y(\bm x)$.
%Assume that $\mathrm{P}(X)$ is the probability density function of $X$, and therefore $\mathrm{P}(X_i \mid X_{- i})$ presents the conditional probability density function of the $i$th component $X_i$ given all the other variables $X_{-i} = (X_1,\dots,X_{i-1},X_{i+1},\dots,X_d)$, and $\mathrm{P}(X_{-i} \mid X_i)$ is the conditional probability density function of $X_{-i}$ given $X_i$. Then $ \mathrm{E}\bigl[Y \mid X_i\bigr] $ denotes the expectation of $Y$ with respect to the conditional distribution of $Y$ given $X_i$, and $\mathrm{Var}\bigl( \mathrm{E}\bigl[ Y \mid X_i\bigl] \bigl)$ represents the variance of the conditional expectation of $Y$, given the random variable $X_i$. Similarly, $\mathrm{Var} \bigl( Y \mid X_{-i} \bigl)$  is the conditional variance of $Y$ given all the variables except the variable $X_i$, and $\mathrm{E} \bigl[ \mathrm{Var} \bigl( Y \mid X_{-i} \bigl) \bigl]$ is the average conditional variance of $Y$ when conditioning on all variables except $X_i$, averaging over the distribution of all the other variables.

Those sensitivity indices are usually computed based on the independent uniform distributions in each dimension of the decision variable, i.e., 
$p(x_1, \ldots, x_d)=\prod^d_{i=1} p_i(x_i)$, where $p_i$ is the probability density function of the uniform distribution over the range of $x_i$. With the decision uncertainty sample in \eqref{eq:mcsample}, we can compute $S_i$ and $S_{T_i}$ with expectations taken with respect to the empirical distribution of \eqref{eq:mcsample}. Therefore, rather than evaluating the impact of input uncertainty using a uniform distribution, we can alternatively assess the sensitivity of the optimal value to the uncertainty estimation of each optimal decision variable. 
By incorporating decision uncertainty estimation into sensitivity analysis, the resulting sensitivity indices capture how the uncertainty in each optimal decision variable contributes to output uncertainty, helping practitioners identify the most critical parameters in the optimization problem.
Our implementation to approximate the sensitivity indices is based on the R Package ``sobolSalt" \citep{Gilquin2016sobolSalt}. If the true function $y(\bm x)$ is not available, the surrogate Gaussian process model can be used in the approximation.

%In our computer experiments, R Package "sobolSalt" \cite{Gilquin2016sobolSalt, RStudio2024} is employed to estimate Sobol' first-order indices and total sensitivity indices. Sensitivity analysis experiments are designed as follows. First, we figure out the Sobol' first-order indices and total sensitivity indices using randomly uniform distributed samples of size $2000$. Second, we employ our proposed approach, using a GP model as a surrogate to approximate the distribution of the optimal solution. $500$ samples are drawn from this distribution, and its Sobol' first-order indices and total sensitivity indices are subsequently estimated.

\begin{figure}[!ht]
    \centering
    \includegraphics[scale=0.2]{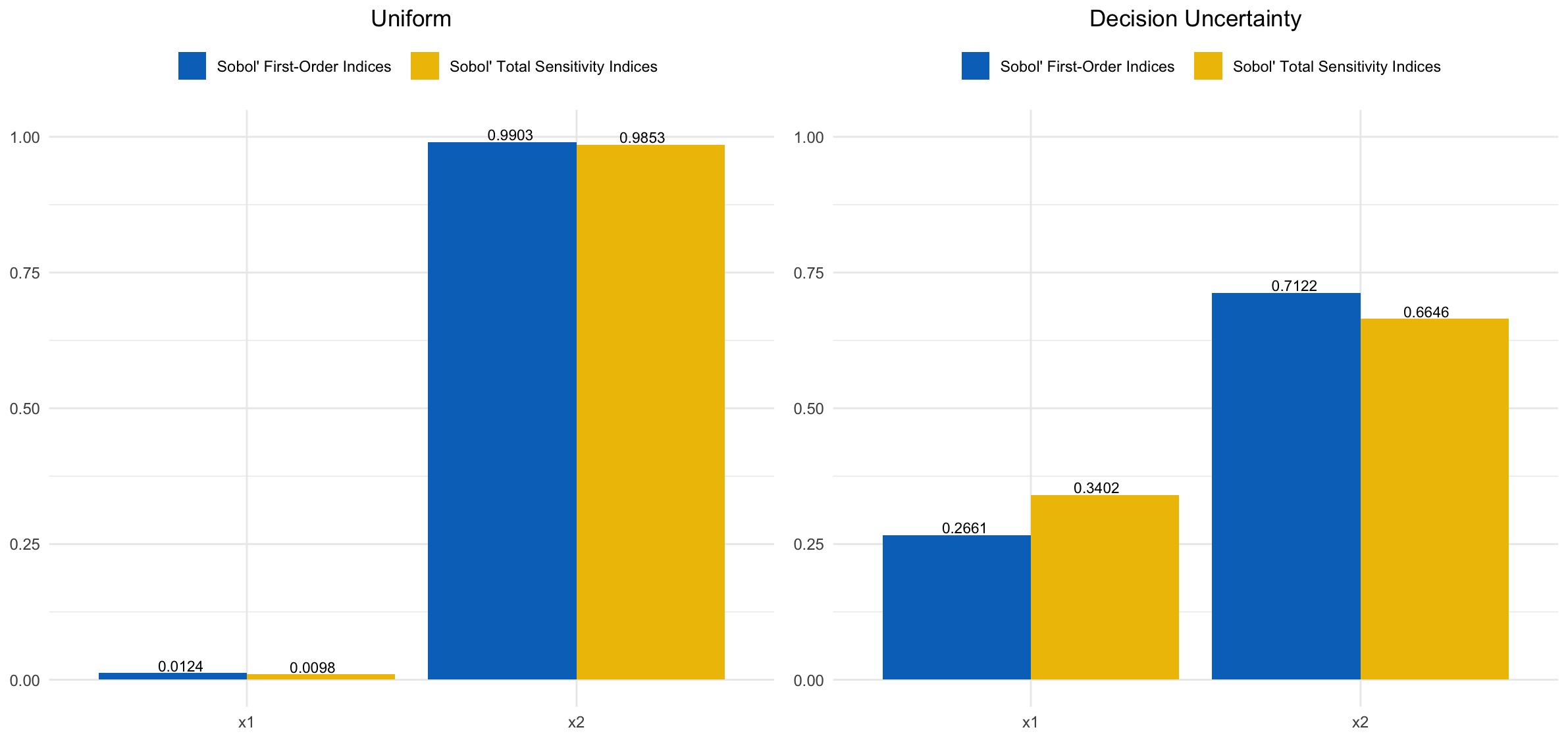}
    \caption{Comparison of sensitivity analysis between the uniform distribution (left) and 
    the decision uncertainty (right) given by \eqref{eq:mcsample}  for the function in \eqref{eq:2stage}.}    \label{fig:sensitivity_analysis_compare}
\end{figure}

We revisit the function in \eqref{eq:2stage} to illustrate the difference between the sensitivity indices calculated based on independent uniform distributions and the empirical distribution in \eqref{eq:mcsample} of optimal decisions. 
Figure \ref{fig:sensitivity_analysis_compare} compares the Sobol first-order indices (blue) and the total sensitivity indices (yellow) for these two scenarios.
For the sensitivity indices computed from the independent uniform distributions of the input variables, the majority of the uncertainty contribution is driven by $x_2$, which matches the expression of the output function in \eqref{eq:2stage}. In contrast, using the decision uncertainty illustrated in Figure \ref{fig:sub2}, the uncertainty contributions from $x_1$ and $x_2$ are approximately divided in a 1:2 ratio, with $x_1$ contributing about $1/3$ and $x_2$ about $2/3$. The uncertainty contribution has incorporated our belief about the optimal decision given by the training dataset.

\section{Application to Cure Process Optimization} \label{sec:case_study}
Thermoset-based fiber-reinforced composite laminates are widely used to manufacture structural components, with the cure process being a key step in their production \citep{mcilhagger2020manufacturing, limaye2025numericalsimulationinformedrapid}. This process involves applying controlled temperature cycles, commonly referred to as cure cycles, to initiate and complete the polymerization reactions of the thermoset prepreg. However, the curing process often induces residual stresses, arising from both internal factors, such as material shrinkage, and external factors, such as uneven temperature distribution. These residual stresses can result in deformations %(show in Figure \ref{fig:deformation}) 
that compromise the structural integrity and dimensional accuracy of the laminate \citep{white1993cure, agius2016rapidly}. Therefore, minimizing or eliminating these process-induced stresses and deformations is essential for manufacturing high-quality composite structures with reliable performance. 

%The existing optimization methods, however, face challenges in scaling to larger, more complex structures, as they often require extensive material characterization and incur high computational costs. To address these challenges, this work proposes Constrained Bayesian Optimization using a Gaussian Process (GP) model as a surrogate. This approach aims to accelerate the optimization process, ensuring efficient and robust manufacturing of advanced composite structures.
%\begin{figure}
%    \centering
%    \includegraphics[scale=1]{deformation.png}
%    \caption{Schematic Diagram of Deformation in Case Study (described in \cite{limaye2025numericalsimulationinformedrapid})}
%    \label{fig:deformation}
%\end{figure}

\begin{figure}
    \centering
    \includegraphics[scale=0.6]{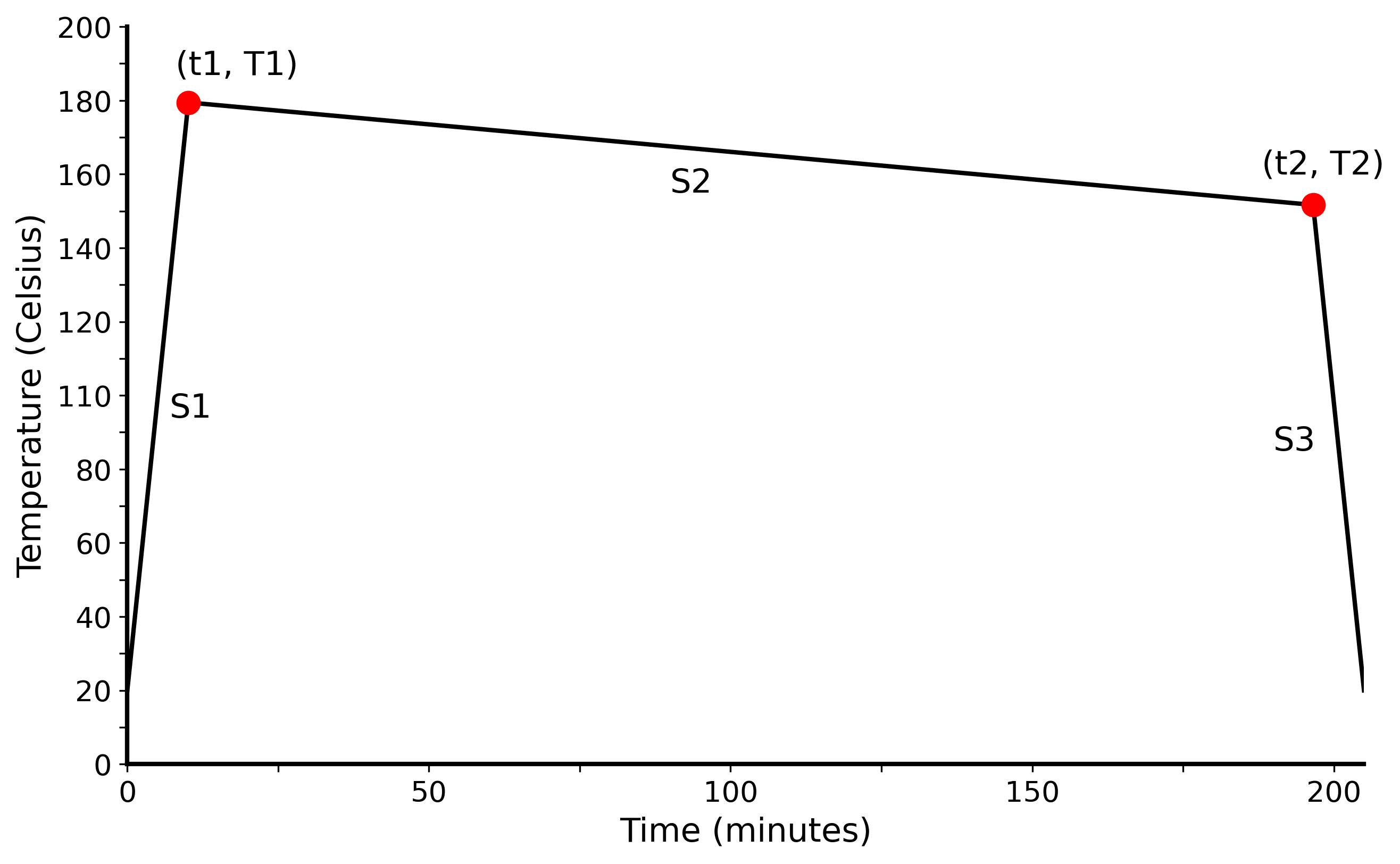}
    \caption{A cure cycle with two change points.}
    \label{fig:4d}
\end{figure}

The cure process can be simulated using a commercial finite element analysis software ABAQUS \citep{abaqus2024}, in conjunction with the cure simulation tool COMPRO \citep{compro2024}. 
As identified by
\cite{limaye2025numericalsimulationinformedrapid},
two specific change points in the cure cycle play a critical role in the final part quality. 
Accordingly, this study simplifies the optimization problem to determine the optimal locations of these two change points, denoted as $(t_1, T_1)$ and $(t_2, T_2)$ as in Figure~\ref{fig:4d}.
This simulation outputs the deformation $y(t_1, T_1, t_2, T_2)$ and the degree of cure $z(t_1, T_1, t_2, T_2)$. The objective is to minimize the deformation while ensuring that the degree of cure is at least 0.96.
As a result, the cure process optimization problem is formulated by
\begin{align} 
    \mbox{min } & y(t_1,T_1,t_2,T_2)  \nonumber\\
    \mbox{s.t. } & z(t_1,T_1,t_2,T_2) \geq 0.96 \label{eq:cureopt} \\
    %& \frac{T_2-T_1}{t_2-t_1} - \frac{T_1-20}{t_1} \le 0 \label{eq:s1} \\
    & \frac{T_1-20}{t_1} \ge 0 \label{eq:s1} \\
    & \frac{T_1-T_2}{t_2-t_1} \le 0 \label{eq:s2} \\
    & 10 \le t_1 \le 110  \nonumber\\
    & 125 \le T_1 \le 180 \nonumber\\
    & 120 \le t_2 \le 200 \nonumber\\
    & 150 \le T_2 \le 180 \nonumber,
\end{align}
where the constraints in \eqref{eq:s1} and 
\eqref{eq:s2} are 
given by requiring that the slope of $S_1$ is greater than or equal to zero, and the slope of $S_2$ is less than or equal to zero as shown in Figure \ref{fig:4d}.

%The design variables for the optimization problem were generated by controlling two points in the time-temperature (t-T) plane along the cure cycle (present in Figure \ref{fig:4d}). The objective of the optimization was to minimize objective function, representing the cure-induced deformation. A constraint was imposed, requiring the degree of cure (DoC) of the composite laminate to exceed $0.96$. Points $(t_1, T_1)$ and $(t_2, T_2)$ are the optimal locations supposed to obtain. $S_1$, $S_2$ and $S_3$ denote slopes in cure process of each phase, and additional slope constraints are imposed $S_1 \ge S_2$ and $S_2 \ge 0$ which can be written in terms of $t_1$, $T_1$, $t_2$ and $T_2$ as $\frac{T_2-T_1}{t_2-t_1} - \frac{T_1-20}{t_1} \le 0$ and $\frac{T_1-T_2}{t_2-t_1} \le 0$ . 
%Therefore, the optimization problem can be formulated in a 4 dimensional case that
%\begin{align} 
%    \mbox{min } & y(t_1,T_1,t_2,T_2) \nonumber \\
%    \mbox{st } & z(t_1,T_1,t_2,T_2) \geq 0.96 \nonumber \\
%    & \frac{T_2-T_1}{t_2-t_1} - \frac{T_1-20}{t_1} \le 0 \nonumber \\
%    & \frac{T_1-T_2}{t_2-t_1} \le 0 \nonumber \\
%    & 10 \le t_1 \le 110 \label{eq:case_study_4d} \\
%    & 125 \le T_1 \le 180 \nonumber\\
%    & 120 \le t_2 \le 200 \nonumber\\
%    & 150 \le T_2 \le 180 \nonumber 
%\end{align}
%where $f(t_1,T_1,t_2,T_2)$ is the objective function for deformation, and $g(t_1,T_1,t_2,T_2)$ is the constraint function from DoC. $(t_1, T_1)$ and $(t_2, T_2)$ are the time and temperature locations splitting the cure cycle into three phases shown in Figure \ref{fig:4d}. 

\begin{figure}[h]
  \centering
\includegraphics[width=1.125\linewidth]{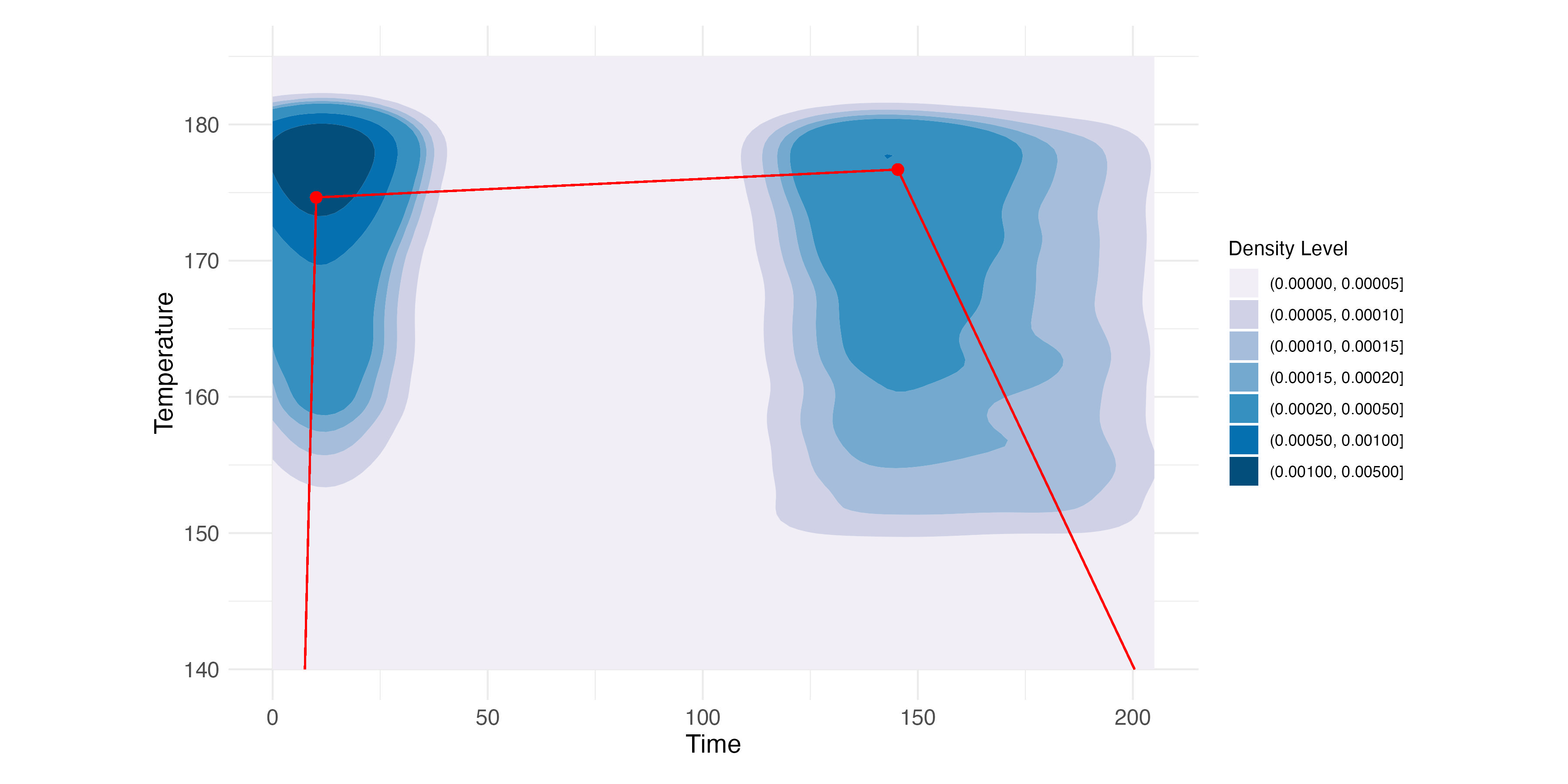}
    \caption{The probability density of
  the empirical distributions of the optimal change points  for the optimization problem in \eqref{eq:cureopt}. The red dots are the optimal value given by the mean functions of the GPs in \eqref{eq:meanopt}.}
    \label{fig:uq1}
\end{figure}

We consider the cure process of an L-shaped laminate in \cite{limaye2025numericalsimulationinformedrapid}. Generate a dataset with $n=50$ runs from the cure process simulation. The data contains four inputs $t_1, T_1, t_2, T_2$ as shown in Figure \ref{fig:4d}, and two outputs $y(t_1, T_1, t_2, T_2)$ and $z(t_1, T_1, t_2, T_2)$. Since this problem has an additional black-box constraint, we modify the uncertainty estimation procedure in Section \ref{sec:gpm} by adding an independent GP surrogate model for $z(t_1, T_1, t_2, T_2)$ with detail provided in the Appendix. We provide the uncertainty estimation results in Figure \ref{fig:uq1}. As in \eqref{eq:mcsample}, we can obtain a sample of optimal points based on the realizations generated by the fitted GPs. Figure \ref{fig:uq1} shows the contour plot of the probability density corresponding to the empirical distributions given by the sample of the optimal points for each of the two change points. This figure shows that, the second change point has higher uncertainty than the second change point based on the fitted GP surrogates.

\begin{figure}
    \centering
    \includegraphics[scale=0.096]{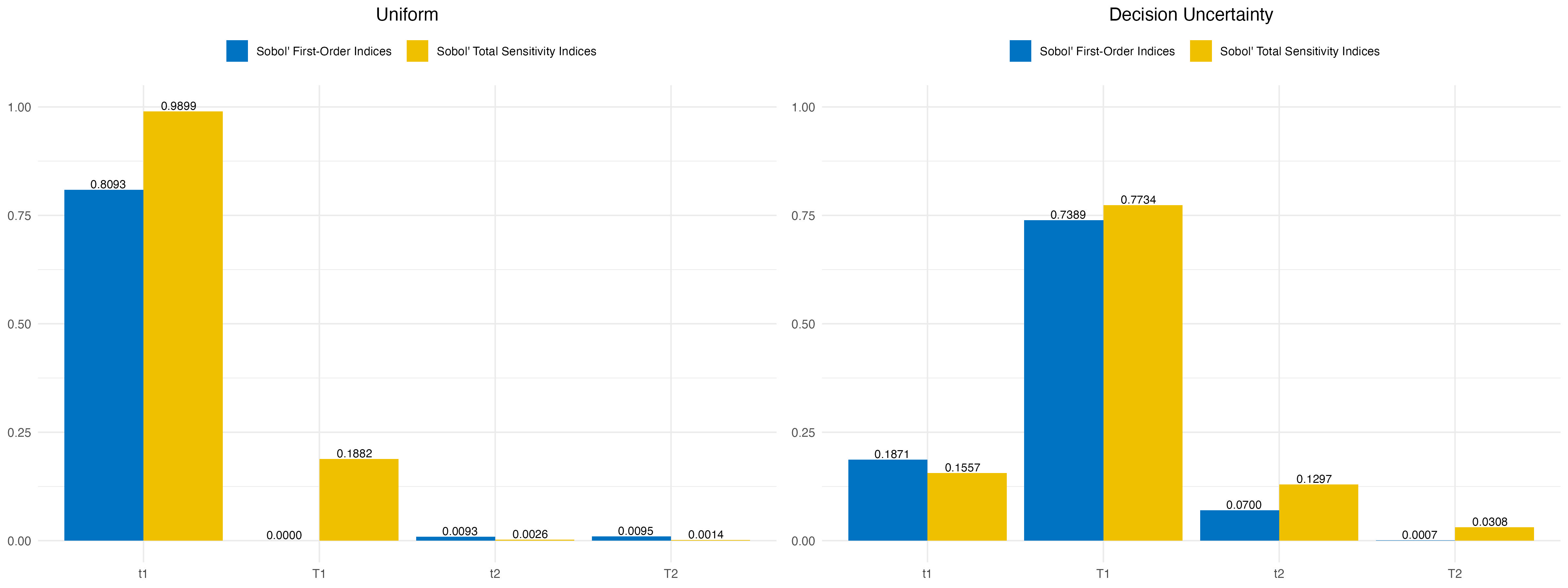}
    \caption{Comparison of sensitivity analysis between the uniform distribution (left) and the decision uncertainty (right) for the optimization problem in \eqref{eq:cureopt}.}
    \label{fig:case_study_sa}
\end{figure}

We also provide sensitivity analysis results using the decision uncertainty in Figure \eqref{fig:uq1}. As in Section \ref{sec:sensitivity_analysis}, we compare the sensitive analysis results given by decision uncertainty with independent uniform distributions of the four inputs. 
The results for the sensitivity indices are provided in 
Figure \ref{fig:case_study_sa}. We have the following findings from this figure. First, based on both uniform input uncertainty and decision uncertainty, the uncertainty of the first change point has more contribution to the output uncertainty than that of the second change point. Second, given the decision uncertainty, the temperature $T_1$ becomes more important than the time $t_1$ for the first change point, whereas the time $t_2$ becomes more important than the temperature $T_2$ for the second change point. From a practical perspective, this result suggests follow-up investigation on the location of the first change point, especially regarding the temperature.

\section{Conclusion} \label{sec:conclusion}

In this work, we developed a surrogate model-based framework for estimating the uncertainty of optimal decisions in black-box optimization problems and assessing the sensitivity of this uncertainty with respect to the objective function. By leveraging the probabilistic nature of Gaussian process models, our method provides an approach to estimate and analyze decision uncertainty. We demonstrated the application of this framework in the context of the composite cure process simulation, where identifying reliable parameter settings is crucial for minimizing product deformation.

\appendix 
\section*{Appendix: Decision Uncertainty Estimation for Optimization Problem with Blackbox Constraints}
In many applications, the optimization problem may also contain blackbox constraints (e.g., \cite{limaye2025numericalsimulationinformedrapid}). We can formulate the problem to be
\begin{equation}
\label{eq:opt:constrained}
\min y(\bm x)
\quad\text{s.t.}\quad
z(\bm x) \ge  c\quad\mathrm{and}\quad \bm x\in\mathcal X,
\end{equation}
where $z(\bm x)$ is a blackbox function and $c$ is a known constant.

Let $\bm z  = \left( z \left( \bm x_1 \right), \cdots, z \left( \bm x_n \right)\right)^\top$. 
We assume that the constraint function $z(\bm x)$ is a realization of a GP independent with $Y(\bm x)$ in \eqref{eq: model1}:
\begin{equation}
    Z \left( \bm x\right) = \mu_z + \epsilon_z \left( \bm x \right), \quad \epsilon_z \left( \bm x \right) \sim \mathcal{GP}\left( 0, \sigma_z^2 r_z(\cdot, \cdot) \right)
    \label{eq: model2}
\end{equation}
with constant mean $\mu_z$, variance $\sigma_z^2$, and the correlation structure $r_z(\cdot, \cdot)$ 
can be different from \eqref{eq:corr}.
%same as $r(\cdot, \cdot)$ in Equation \eqref{eq: model1} but with different correlation parameters,  
%\[
%r_z(\bm x_i, \bm x_j)= \mathrm{Corr}\left[ \epsilon_z \left( \bm x_i \right), \epsilon_z \left( \bm x_j \right) \right].
%\]
Following the conditional GP of $Y(\bm x)$ in \eqref{postGP:sigma} in Section \ref{sec:gpm}, 
we have that
\begin{equation}\label{postGP:sigma_z}
 Z \left( \bm x \right) \mid \bm z\sim \mathcal{GP} \left( \hat{z} \left( \bm x \right), \Sigma_z \left( \bm x, \bm x'\right) \right),
 \end{equation}
 where 
\[
\hat{z} \left( \bm x \right) = \hat{\mu}_z +\bm r_z(\bm x)^\top \bm R^{-1}_z \left(\bm z- \bm 1 \hat{\mu}_z \right)
\]
and \[
\Sigma_z(\bm x,\bm x^{\prime})
=
\sigma^{2}_z\!\left[
\,r_z(\bm x,\bm x^{\prime})\;-\;
\bm r_z(\bm x)^{\!\top}\bm R_z^{-1}\bm r_z(\bm x^{\prime})
\;+\;
\frac{\Bigl(1-\mathbf 1^{\!\top}\bm R_z^{-1}\bm r_z(\bm x)\Bigr)
\Bigl(1-\mathbf 1^{\!\top}\bm R_z^{-1}\bm r_z(\bm x^{\prime})\Bigr)}
{\mathbf 1^{\!\top}\bm R_z^{-1}\mathbf 1} \right]
\,
\] 
with $\bm R_z$ being an $n \times n$ matrix with the $\left( i, j \right)$-th entry $r_z \left(\bm x_i, \bm x_j \right)$, and
\[\bm r_z(\bm x)=
\begin{bmatrix}
r_z(\bm x,\bm x_1), 
r_z(\bm x,\bm x_2),
\ldots,
r_z(\bm x,\bm x_n)
\end{bmatrix}^\top
\quad \mathrm{and}\quad
    \hat{\mu}_z = \frac{\bm 1^\top \bm R_z ^{-1} \bm z}{\bm 1^\top \bm R_z ^{-1} \bm 1}.
\]

Based on the conditional GPs in \eqref{postGP:sigma} and
\eqref{postGP:sigma_z}, an estimation of the optimal decision to \eqref{eq:opt:constrained} is given by
\begin{equation}\label{eq:meanopt}
\hat{\bm x}^\ast\in \arg \min_{\substack{{\bm x \in \mathcal{X}}}: \hat{z}(\bm x) \ge c} \hat y\left( \bm x \right).
\end{equation}
To generate $M$ samples of the optimal decision $\bm x^\ast$, we draw $M$ independent realizations
from the posterior Gaussian process $Y \left( \bm x \right) \mid \bm y$ in \eqref{postGP:sigma} and, for each realization, identify the optimal solution within the feasible region defined by the posterior Gaussian process $Z \left( \bm x \right) \mid \bm z$ in \eqref{postGP:sigma_z}. Let $\tilde y_{(i)}(\bm x)$ and $\tilde z_{(i)}(\bm x)$ with $\bm x\in\mathcal X$ be the $i$-th realization of the conditional Gaussian processes $Y \left( \bm x \right) \mid \bm y$ and $Z \left( \bm x \right) \mid \bm z$, respectively. Then, for $i=1, \ldots, M$, we obtain
\[
\bm x^\ast_{(i)}\in \arg \min_{\substack{{\bm x \in \mathcal{X}_{(i)}}} : \tilde z _{(i)}(\bm x) \ge c} \tilde y_{(i)}(\bm x)
\]
And the ensemble of optimal points
$\{\bm x^\ast_{(1)}, \dots, \bm x^\ast_{(M)}\}$ forms a random sample for quantifying the uncertainty in the constrained optimal decision.

In our implementation, we sample the realizations of Gaussian processes $\tilde y_{(i)}(\bm x)$ and $\tilde z_{(i)}(\bm x)$ on a finite set of input values. First, for each realization $i$, we generate an independent $d$-dimensional Latin hypercube of $N$ runs, denoted $\mathcal X_{(i),N}=\{\bm x_{(i),1}, \ldots, \bm x_{(i), N}\}$. Subsequently,
we obtain the mean vectors $(\hat y(\bm x_{(i),1}), 
\ldots, \hat y(\bm x_{(i),N}))^\top$ and $(\hat z(\bm x_{(i),1}), 
\ldots, \hat z(\bm x_{(i),N}))^\top$, and the $N\times N$ covariance matrices with $(jk)$-th entry $\Sigma(\bm x_{(i),j}, \bm x_{(i),k})$ and $\Sigma_z(\bm x_{(i),j}, \bm x_{(i),k})$, corresponding to the Gaussian process models for objective and constraint functions respectively. Then we generate random vectors
\[(\tilde y(\bm x_{(i),1}), 
\ldots, \tilde y(\bm x_{(i),N}))^\top \quad \text{and} \quad (\tilde z(\bm x_{(i),1}), 
\ldots, \tilde z(\bm x_{(i),N}))^\top\]
from each of the two multivariate normal distributions.
 Finally, we collect the random sample of the optimal point over $M$ replications
 \[
\{ \bm x^\ast_{(1),N},\ldots, \bm x^\ast_{(M),N}\}
\] 
with 
\[
\bm x^\ast_{(i),N}\in \arg \min_{\substack{{\bm x \in \mathcal{X}_{(i),N}}} : \tilde z _{(i)}(\bm x) \ge c} \tilde y_{(i)}(\bm x).
\]
For the application to cure process optimization in Section \ref{sec:case_study}, we use $M=10000$ and $N=500$.

\section*{Acknowledgment}
This work is based in part on the first author's doctoral dissertation  at Clemson University. 
This work was supported as part of the AIM for Composites, an Energy Frontier Research Center funded by the U.S. Department of Energy, Office of Science, Basic Energy Sciences at the Clemson University under award \#DE-SC0023389.

\section*{Data Availability Statement}
The data that support the findings of this study are openly available in Github at \url{https://github.com/yezhuoli/Uncertainty-Estimation-of-the-Optimal-Decision}. 

\bibliographystyle{asa}   
\bibliography{ref} 

\begin{thebibliography}{18}
\newcommand{\enquote}[1]{``#1''}
\expandafter\ifx\csname natexlab\endcsname\relax\def\natexlab#1{#1}\fi

\bibitem[{Agius et~al.(2016)Agius, Joosten, Trippit, Wang, and Hilditch}]{agius2016rapidly}
Agius, S., Joosten, M., Trippit, B., Wang, C., and Hilditch, T. (2016), \enquote{Rapidly cured epoxy/anhydride composites: Effect of residual stress on laminate shear strength,} \textit{Composites Part A: Applied Science and Manufacturing}, 90, 125--136.

\bibitem[{{Convergent Manufacturing Technologies, Inc.}(2024)}]{compro2024}
{Convergent Manufacturing Technologies, Inc.} (2024), \textit{COMPRO – Composite Process Modeling Software}, Vancouver, BC, computer software.

\bibitem[{{Dassault Systèmes}(2024)}]{abaqus2024}
{Dassault Systèmes} (2024), \textit{{ABAQUS 2024 User Manual}}, Dassault Systèmes Simulia Corp., Providence, RI.

\bibitem[{Frazier(2018)}]{frazier2018tutorial}
Frazier, P.~I. (2018), \enquote{A tutorial on Bayesian optimization,} \textit{arXiv preprint arXiv:1807.02811}.

\bibitem[{Gilquin(2016)}]{Gilquin2016sobolSalt}
Gilquin, L. (2016), \textit{sobolSalt: Monte Carlo Estimation of Sobol' Indices Based on Saltelli's Schemes}, r package version 1.30.1.

\bibitem[{Gramacy(2020)}]{gramacy2020surrogates}
Gramacy, R.~B. (2020), \textit{Surrogates: Gaussian process modeling, design, and optimization for the applied sciences}, Chapman and Hall/CRC.

\bibitem[{Jones et~al.(1998)Jones, Schonlau, and Welch}]{jones1998efficient}
Jones, D.~R., Schonlau, M., and Welch, W.~J. (1998), \enquote{Efficient global optimization of expensive black-box functions,} \textit{Journal of Global optimization}, 13, 455--492.

\bibitem[{Limaye et~al.(2025)Limaye, Li, Zhang, and Li}]{limaye2025numericalsimulationinformedrapid}
Limaye, M., Li, Y., Zhang, Q., and Li, G. (2025), \enquote{Numerical Simulation Informed Rapid Cure Process Optimization of Composite Structures using Constrained Bayesian Optimization,} .

\bibitem[{Marrel et~al.(2009)Marrel, Iooss, Laurent, and Roustant}]{marrel2009calculations}
Marrel, A., Iooss, B., Laurent, B., and Roustant, O. (2009), \enquote{Calculations of Sobol indices for the Gaussian process metamodel,} \textit{Reliability Engineering \& System Safety}, 94, 742--751.

\bibitem[{McIlhagger et~al.(2020)McIlhagger, Archer, and McIlhagger}]{mcilhagger2020manufacturing}
McIlhagger, A., Archer, E., and McIlhagger, R. (2020), \enquote{Manufacturing processes for composite materials and components for aerospace applications,} in \textit{Polymer composites in the aerospace industry}, Elsevier, pp. 59--81.

\bibitem[{McKay(1992)}]{mckay1992latin}
McKay, M.~D. (1992), \enquote{Latin hypercube sampling as a tool in uncertainty analysis of computer models,} in \textit{Proceedings of the 24th conference on Winter simulation}, pp. 557--564.

\bibitem[{Roustant et~al.(2012)Roustant, Ginsbourger, and Deville}]{roustant2012dicekriging}
Roustant, O., Ginsbourger, D., and Deville, Y. (2012), \enquote{DiceKriging, DiceOptim: Two R packages for the analysis of computer experiments by kriging-based metamodeling and optimization,} \textit{Journal of statistical software}, 51, 1--55.

\bibitem[{Sacks et~al.(1989)Sacks, Welch, Mitchell, and Wynn}]{sacks1989statistical}
Sacks, J., Welch, W., Mitchell, T., and Wynn, H. (1989), \enquote{Statistical science,} \textit{Design and Analysis of Computer Experiments}, 4, 409--423.

\bibitem[{Santner et~al.(2003)Santner, Williams, Notz, and Williams}]{santner2003design}
Santner, T.~J., Williams, B.~J., Notz, W.~I., and Williams, B.~J. (2003), \textit{The design and analysis of computer experiments}, vol.~1, Springer.

\bibitem[{Storlie and Helton(2008)}]{storlie2008multiple}
Storlie, C.~B. and Helton, J.~C. (2008), \enquote{Multiple predictor smoothing methods for sensitivity analysis: Description of techniques,} \textit{Reliability Engineering \& System Safety}, 93, 28--54.

\bibitem[{White and Hahn(1993)}]{white1993cure}
White, S.~R. and Hahn, H. (1993), \enquote{Cure cycle optimization for the reduction of processing-induced residual stresses in composite materials,} \textit{Journal of Composite Materials}, 27, 1352--1378.

\bibitem[{Williams and Rasmussen(2006)}]{williams2006gaussian}
Williams, C.~K. and Rasmussen, C.~E. (2006), \textit{Gaussian processes for machine learning}, MIT press Cambridge, MA.

\bibitem[{Zhang and Hwang(2020)}]{zhang2020sequential}
Zhang, Q. and Hwang, Y. (2020), \enquote{Sequential model-based optimization for continuous inputs with finite decision space,} \textit{Technometrics}, 62, 486--498.

\end{thebibliography}

\end{document}